\documentstyle[11pt]{article}
\newfont{\goth}{eufm10 scaled \magstep1}

\def\go{\mbox{\goth o}}
\def\gp{\mbox{\goth p}}

\def\gs{\mbox{\goth s}}

\def\gu{\mbox{\goth u}}

\def\a{\alpha}
\def\b{\beta}
\def\c{\gamma}\def\C{\Gamma}
\def\d{\delta}

\def\h{\eta}
\def\k{\kappa}

\def\th{\theta}\def\Th{\Theta}

\def\beq{\begin{equation}}\def\eeq{\end{equation}}
\def\beqa{\begin{eqnarray}}\def\eeqa{\end{eqnarray}}
\def\barr{\begin{array}}\def\earr{\end{array}}

\def\del{\partial}
\def\ua{\underline{\alpha}}
\def\ub{\underline{\phantom{\alpha}}\!\!\!\beta}
\def\uc{\underline{\phantom{\alpha}}\!\!\!\gamma}
\def\ud{\underline\delta}

\def\una{\underline a}\def\unA{\underline A}
\def\unb{\underline b}\def\unB{\underline B}
\def\unc{\underline c}\def\unC{\underline C}
\def\und{\underline d}
\def\une{\underline e}

\def\unM{\underline M}

\def\oa{\overline{\alpha}}\def\ob{\overline{\phantom{\alpha}}\!\!\!\beta}
\def\ova{\overline{a}}
\def\ovA{\overline{A}}

\let\bm=\bibitem

\def\bd{\begin{document}}
\def\ed{\end{document}}
\def\ba{\begin{array}}
\def\ea{\end{array}}
\def\bea{\begin{eqnarray}}
\def\eea{\end{eqnarray}}
\def\ft#1#2{{\textstyle{{\scriptstyle #1}\over {\scriptstyle #2}}}}
\def\fft#1#2{{#1 \over #2}}
\newcommand{\be}{\begin{equation}}
\newcommand{\ee}{\end{equation}}
\newcommand{\eq}[1]{(\ref{#1})}
\def\eqs#1#2{(\ref{#1}-\ref{#2})}

\begin{document}

\begin{titlepage}
\begin{flushright}
King's College/kcl-th-96-16\\
Texas A \& M/CTP TAMU-55/96\\
\hfill{hep-th/9611008}\\
\today
\end{flushright}
\vskip 2cm
\begin{center}
{\Large{\bf $D=11$, $p=5$}}
\end{center}
\vskip 1.5cm
\centerline{\bf P.S. Howe,}
\vskip .5cm
\centerline{Dept. of Mathematics,}
\vskip .5cm
\centerline{King's College,  London, UK.}
\vskip 5mm
\centerline{and}
\vskip 5mm
\centerline{\bf E. Sezgin,}
\vskip .5cm
\centerline{Center for
Theoretical Physics,}
\vskip .5cm
\centerline{Texas A \& M University,}
\vskip .5cm
\centerline{College Station, Texas, US.}
\vskip 1.5cm

\begin{abstract}

\noindent
The equations of motion of the super five-brane in $D=11$ dimensions are
derived using the formalism of superembeddings. The equations describe
highly nonlinear self-interactions of a tensor multiplet in the six
dimensional worlsurface, and they have manifest worldsurface local
supersymmetry. The geometry of the target space corresponds to $D=11$
supergravity. 

\end{abstract}

\end{titlepage}

In a recent paper \cite{hs} it was shown that all super $p$-branes
preserving half-supersymmetry and with $N=1$ target space supersymmetry,
for any dimension $D$ of spacetime, or $N=2$ for $D=10$, can be
understood from the perspective of superembeddings of one supermanifold,
the worldsurface, into another, the target superspace, and furthermore
that the basic superembedding condition which determines the
worldsurface multiplets is both very natural and universal. The analysis
given in \cite{hs} was mainly at the linearised level but is applicable
to all branes whether they are type I or type II, where type II branes
are those which have physical worldsurface bosons which are not all
scalars. Reference \cite{hs} follows in the tradition of the doubly
supersymmetric approach to supersymmetric extended objects initiated in
\cite{stv} for which we refer the reader to \cite{b1}, where the $D=11$ supermembrane was discussed as a supermebedding in a flat target superspace, for a full set
of references to the earlier literature. In this paper we briefly report
on the full non-linear equations of motion for the 5-brane in $D=11$.
Partial results for the bosonic sector have been obtained previously
\cite{pkt1,ah,eric1,witten}, but the superspace approach, as we shall show,
gives a systematic method for determining the full system of equations
of motion. A detailed discussion of both the 2-brane and the 5-brane in
D=11 in arbitrary backgrounds is in preparation.

We consider embeddings $M\hookrightarrow\unM$, where the worldsurface
$M$ has (even$|$odd) dimension $(6|16)$ and the target space, $\unM$,
has dimension $(11|32)$. In local coordinates $z^{\unM}$ for $\unM$ and
$z^M$ for $M$ the embedded submanifold is given as $z^{\unM}(z)$. We
define the embedding matrix $E_A{}^{\unA}$ to be the derivative of the
embedding referred to preferred bases on both manifolds: 
\beq
E_A{}^{\unA}=E_A{}^M\del_{M}z^{\unM}E_{\unM}{}^{\unA},
\eeq
where $E_M{}^A\ (E_A{}^M)$ is the supervielbein (inverse supervielbein)
which relates the preferred frame basis to the coordinate basis, and the
target space supervielbein has underlined indices. The notation is as
follows: indices from the beginning (middle) of the alphabet refer to
frame (coordinate) indices, latin (greek) indices refer to even (odd)
components and capital indices to both, non-underlined (underlined)
indices refer to $M\ (\unM)$ and primed indices refer to normal
directions. We shall also employ a two-step notation for spinor indices;
that is, for general formulae a spinor index $\a$ (or $\a'$) will run
from 1 to 16, but to interpret these formulae we shall replace a
subscript $\a$ by a subscript pair $\a i$ and a subscript $\a'$ by a
pair ${}_i^{\a}$, where $\a=1,\dots 4$ and $i=1,\dots 4$ reflecting the
$Spin(1,5)\times USp(4)$ group structure of the $N=2,d=6$ worldsurface
superspace. (A lower (upper) $\a$ index denotes a left-handed
(right-handed) $d=6$ Weyl spinor and the $d=6$ spinors that occur in the
theory are all symplectic Majorana-Weyl.) 

We shall find it convenient to introduce a basis for the normal
bundle, $E_{A'} =(E_{a'},E_{\a'})$, given in terms of a basis of the
tangent bundle of the target space $E_{\unA}$ (at any point $p\in M$) by
\beq
E_{A'}=E_{A'}{}^{\unA} E_{\unA}\ .
\eeq
We can assemble the embedding matrix and the normal matrix into a square matrix which we shall denote by $E_{\ovA}{}^{\unA}=(E_A{}^{\unA},E_{A'}{}^{\unA})$. The inverse of this matrix will be written as $E_{\unA}{}^{\ovA}=(E_{\unA}{}^{A},E_{\unA}{}^{A'})$

The basic equation describing the embedding is
\beq
E_{\a}{}^{\una}=0\ . \label{mc1}
\eeq
Its geometrical meaning is that the odd tangent space of the
worldsurface is a subspace of the odd tangent space of the target space
at each point $p\in M$. As we shall see later, an immediate consequence
of (\ref{mc1}) is 
\beq
E_\a{}^{\ua}E_\b{}^{\ub}T_{\ua\ub}{}^{\unc}=T_{\a\b}{}^cE_c{}^{\unc}\ .\label{mc2}
\eeq
We can also choose the odd normal tangent bundle (which is not fixed by
the embedding) such that 
\beq
E_{\a'}=E_{\a'}{}^{\ua} E_{\ua}\ , \label{e1}
\eeq
in other words $E_{\a'}{}^{\una}=0$. This in turn implies that
\beq
E_{\ua}{}^a=E_{\ua}{}^{a'}=0\ . \label{e2}
\eeq
As a consequence of \eq{mc1}, \eq{e1} and \eq{e2}, one finds that the inverse $E$ 
in the even-even sector is the inverse of the even-even part of $E$ and
similarly for the odd-odd sector. 

Equations \eq{mc1} and \eq{mc2} are the fundamental equations. We observe that 
they do not require a fixed choice of either the worldsurface or target space even 
tangent bundle, and that no connections are involved. Nevertheless, in order to 
work out the consequences of these equations it is useful to make appropriate 
choices of these objects. We shall assume that the target space supergeometry 
corresponds to on-shell $D=11$ supergravity. The structure group is $Spin(1,10)$. 
All of the components of the torsion vanish except for \cite{cf,bh}
\beq
T_{\ua\ub}{}^{\unc}=-i(\C^{\unc})_{\ua\ub}\ ,
\eeq
\beq
T_{\una\ub}{}^{\uc}=-
{1\over36}(\C^{\unb\unc\und})_{\ub}{}^{\uc}H_{\una\unb\unc\und}
-{1\over288}(\C_{\una\unb\unc\und\une})_{\ub}{}^{\uc}H^{\unb\unc\und\une}\ ,
\eeq
where $H_{\una\unb\unc\und}$ is totally antisymmetric, and the dimension 3/2 
component $T_{\una\unb}{}^{\uc}$. $H_{\una\unb\unc\und}$ is the dimension one 
component of the closed superspace 4-form $H_4$ whose only other non-vanishing 
component is
\beq
H_{\una\unb\uc\ud}=-i(\C_{\una\unb})_{\uc\ud}\ .
\eeq
With these assumptions \eq{mc2} becomes
\beq
E_\a{}^{\ua}E_\b{}^{\ub}(\C^{\unc})_{\ua\ub}=iT_{\a\b}{}^cE_c{}^{\unc}
\label{me}\ .
\eeq

The solution to this equation is given by
\beq
E_{\a}{}^{\ua}=u_{\a}{}^{\ua}+h_{\a}{}^{\b'} u_{\b'}{}^{\ua}\ ,
\eeq
and
\beq
E_{a}{}^{\una}=u_{a}{}^{\una}-72k_a{}^b u_b{}^{\una}\ ,
\eeq
together with
\beq
T_{\a\b}{}^c=-i(\C^c)_{\a\b}\rightarrow -i\h_{ij}(\c^c)_{\a\b}\ .
\eeq
with $\h_{ij}=-\h_{ji}$ being the $USp(4)$ invariant tensor. In
addition, $u_{\a}{}^{\ua}$ and $u_{\a'}{}^{\ua}$ together make up a
$32\times 32$ matrix $u_{\oa}{}^{\ua}$ which is an element of
$Spin(1,10)$ and $u_a{}^{\una},\ u_{a'}{}^{\una}$ together make up the
corresponding element of the Lorentz group which we shall denote by
$u_{\ova}{}^{\una}$, so that 
\beq
u_{\oa}{}^{\ua} u_{\ob}{}^{\ub} (\C^{\una})_{\ua\ub}=
i(\C^{\ova})_{\oa\ob}u_{\ova}{}^{\una}\ .
\eeq
The tensor $h_\a{}^{\b'}$ is given by
\beq
h_{\a}{}^{\b'}\rightarrow h_{\a i \b}^{\phantom{\a i} j}=
\d_i{}^j(\c^{abc})_{\a\b} h_{abc}\ ,
\eeq
where $h_{abc}$ is self-dual. Finally, $k_{ab}$, defined by
\beq
k_{ab}=h_a{}^{cd} h_{bcd}\ ,
\eeq
is symmetric and traceless. Given the above solution one can find
another by acting with the group $Spin(1,5)\times USp(4)$ on the $d=6$
odd indices $\a$ and with the Lorentz group on the vector indices. There
is also the freedom to make Weyl rescalings; since the theory is
invariant under these transformations we can, in particular, take the
conformal factor to be equal to one. 

The odd-odd and even-even components of the normal matrix
$E_{A'}{}^{\unA}$ can be chosen to be 
\beq
E_{\a'}{}^{\ua}=u_{\a'}{}^{\ua}\ ,
\eeq
and
\beq
E_{a'}{}^{\una}=u_{a'}{}^{\una}\ ,
\eeq
and the inverses in the odd-odd and even-even sectors are
\beq
E_{\ua}{}^{\a}=u_{\ua}{}^{\a}\ , \qquad
E_{\ua}{}^{\a'}=u_{\ua}{}^{\a'}-u_{\ua}{}^{\b} h_{\b}{}^{\a'}\ ,
\eeq
and
\beq
E_{\una}{}^a=u_{\una}{}^b(m^{-1})_b{}^a\ ,\qquad E_{\una}{}^{a'}=u_{\una}{}^{a'}\ ,
\eeq
where we have introduced the inverses of the group matrices $u$ with
similar index conventions, and where 
\beq
m_a{}^b=\d_a{}^b-72 k_a{}^b \label{m}\ .
\eeq 
We will assume that the matrix $m$ is invertible. The special
configurations of $h$ for which ${\rm det}\, m $ vanishes require special care.
Such singular points in the field space presumably correspond to a
new kind of phase transition. This point deserves further study, and we
hope to address it in the future. It may be related to the singular
points in field space which arise in the context of the
Dirac-Born-Infeld action.

The above results show how the odd tangent spaces of the worldsurface
are related to the odd tangent spaces of the target space. However, the
even tangent space of the world surface is not fixed. We can choose it,
and the worldsurface connection (which takes its values in the Lie
algebra $\gs\go(1,5)\oplus\gu\gs\gp(4)$), so that the torsion
constraints on the worldsurface take a convenient form. In this instance
they turn out to be the constraints of $N=2,d=6$ conformal supergravity.
In what follows we shall not need all the details of this, but we note
that 
\beq
T_{\a b}{}^{c}=T_{\a\b}{}^{\c}=T_{ab}{}^c=0\ .
\eeq

The consequences of the embedding equations can now be analysed
systematically by going through a set of identities which arises from
pulling back the defining equation of the target space torsion 2-form to
the worldsurface. These are 
\beq
\nabla_A E_B{}^{\unC}-(-1)^{AB}\nabla_{B} E_{A}{}^{\unC} +T_{AB}{}^C E_C{}^{\unC}
=(-1)^{A(B+\unB)}E_B{}^{\unB} E_A{}^{\unA} T_{\unA\unB}{}^{\unC}\ , \label{de}
\eeq
where $\nabla$ is covariant with respect to both the worldsurface and
target space structure groups. The dimension zero component of \eq{de} is
simply \eq{mc2}. We shall not give all the details of the analysis of the
rest of these equations here but restrict our attention to indicating
how the equations of motion arise. At dimension one-half one finds 
\beq
\nabla_{\a} E_{\b}{}^{\uc} +\nabla_{\b} E_{\a}{}^{\uc}=i(\C^c)_{\a\b}E_c{}^{\c}\ .
\label{dim1/2}
\eeq
Defining
\beq
\chi_a{}^{\a'}=E_a{}^{\ua} E_{\ua}{}^{\a'}\ ,
\eeq
one finds, by right-multiplying equation \eq{dim1/2} by $E_{\ua}{}^{\a'}$, the
Dirac equation for the spin one-half fermions 
\beq
(\c^a)^{\a\b}\chi_{a\b}^{\phantom{a}j}=0\ .\label{1}
\eeq
At dimension one one has
\beq
\nabla_a E_\b{}^{\uc}-\nabla_\b E_a{}^{\uc} +T_{a\b}{}^{\d} E_{\d}{}^{\uc}=
E_{\b}{}^{\ub} E_{a}^{\una} T_{\una\ub}{}^{\unc}\ .\label{dim1}
\eeq
Right-multiplying by $E_{\uc}{}^{\c'}$, one observes that the term
involving the worldsurface torsion drops out, and that by using the
Dirac equation one can derive 
\beq
\h^{ab} K_{a b}{}^{c'}={1\over8}(\c^{c'})^{jk}(\c^a)^{\b\c}Z_{a\b j\c k}\ ,
\label{dilaton}
\eeq
and
\beq
\hat\nabla^c h_{abc}=-{\h^{jk}\over96}\big( (\c_{[a})^{\b\c}Z_{b],\b\c,jk}
+{1\over2}(\c_{ab}{}^c)^{\b\c}Z_{c\b j\c k}\big)\ , \label{heq}
\eeq 
where
\beq
Z_{a\b}{}^{\c'}=E_{\b}{}^{\ub} E_a{}^{\una} T_{\una\ub}{}^{\uc}E_{\uc}{}^{\c'}
-E_a{}^{\uc}\nabla_{\b} E_{\uc}{}^{\c'}\ .
\eeq
The first term in $Z$ is therefore simply a projection of the dimension
one torsion; the second term involves the product of a dimension 1/2 $E$
with the odd derivative of a dimension zero $E$. Both of these
quantities are determined from the dimension 1/2 equations, and all
dimension 1/2 quantities are expressible in terms of $\chi$. This term
is therefore bilinear in $\chi$ but also has a somewhat complicated
dependence on $h_{abc}$. The hatted covariant derivative in \eq{heq} is
defined as follows: 
\beq
\hat\nabla_a h_{bcd}=\nabla_a h_{bcd}-3X_{a,[b}{}^e h_{cd]e}\ ,
\eeq
with
\beq
X_{a,b}{}^c=(\nabla_a u_b{}^{\unc}) u_{\unc}{}^{c}\ .
\eeq
The left-hand side of \eq{dilaton} is part of the second fundamental form
of the surface which we define to be 
\beq
K_{AB}{}^{C'}=(\nabla_A E_B{}^{\unC})E_{\unC}{}^{C'}\ .
\eeq 
The spinor, scalar and tensor equations of motion are the leading
components in the worldsurface $\th$-expansions of equations \eq{1},\eq{dilaton}
and \eq{heq}, respectively. To see that this identification is correct we can
appeal to the linearised case where 
\beqa
\chi_{a}{}^{\c'}&\rightarrow&\del_a \Th^{\c'}\ ,\\
K_{ab}{}^{c'}&\rightarrow&\del_a\del_b X^{c'}\ ,
\eeqa
in a physical gauge, $X^{a'}$ and $\Th^{\a'}$ being the transverse
coordinate superfields which describe the excitations of the brane.
(These are not independent superfields however, because of \eq{mc2}, see
\cite{hs}.)

Since $h_{abc}$ is self-dual, \eq{heq} also implies a modified bosonic
Bianchi identity for the tensor fields. However, in order to introduce a
corresponding 2-form potential it is necessary to find a superspace
3-form $H_3$ which obeys a 4-form Bianchi identity. From earlier results
\cite{pkt1,ah,eric1}, and by comparison with Dirichlet-branes in $D=10$
\cite{polch1}, we know that the identity we should expect to hold should
have the form 
\beq
dH_3=-\ft1{24} H_4\ ,\label{bi}
\eeq
where $H_4$ is the target space 4-form pulled back onto the
worldsurface. It follows that $H_3$ can be  written locally as
\be
H_3=dB_2-\ft1{24} C_3\ , 
\ee
where $B_2$ is a super two-form potential on the worldsurface, and $C_3$ is the
pullback of the target space super three-form such that $H_4=dC_3$.

The Bianchi identity \eq{bi} was verified at the linearised level in \cite{hs}. The
claim is that \eq{bi} is indeed satisfied provided that the only
non-vanishing component of $H_3$ is the one with purely even indices,
$H_{abc}$. To prove this, one has to systematically check the various
components of the Bianchi identity \eq{bi}. Since most of the components of
$H_3$ and $H_4$ vanish many of these equations are trivially satisfied.
The first non-trivial equation arises at dimension zero: 
\beq
-24(\C^c)_{\a\b} H_{abc}=E_{\a}{}^{\ua} E_{\b}{}^{\ub}E_{a}{}^{\una} 
E_b{}^{\unb}(\C_{\una\unb})_{\a\b}\ .\label{bi2}
\eeq
It is not obvious that the right-hand side of \eq{bi2} has the same
structure as the left-hand side, but it is nevertheless true as one can
show directly using \eq{me} and the `membrane' identity
\beq
(\C^{\una})_{(\ua\ub} (\C_{\una\unb})_{\uc\ud)}=0\ .
\eeq
Using \eq{bi2} one finds the relation between $h$ and $H$; it is
\beq
H_{abc}=h_{abc}-2\cdot 72\, k_a{}^d h_{bcd} +(72)^2 k_a{}^d k_b{}^e h_{cde}\ .
\label{H}
\eeq
This can be rewritten as
\beq
H_{abc}=m_a{}^d m_b{}^e h_{cde}.
\eeq
It is easy to check that the second and third terms in \eq{H} are indeed
antisymmetric and that the second term is anti-self-dual while the third
term is self-dual. Thus the actual field strength tensor $H_{abc}$ is
not itself self-dual but is determined by its self-dual part. The
dimension 1/2 Bianchi identity determines the variation of $H_{abc}$,
$\nabla_{\a} H_{abc}$, in terms of known quantities, i.e. in terms of
$\chi$ and $h$, while the dimension one identity is the $x$-space
Bianchi identity in covariantised form. In view of the modified
self-duality satisfied by $H_{abc}$ this can be viewed as the equation
of motion for the tensor field written in terms of $H_{abc}$ rather than
$h_{abc}$. We emphasise the fact that the Bianchi identity \eq{bi} is 
satisfied automatically provided that we define the components of $H_3$ as 
above; it does not contain any new information but enables us to deduce 
more easily the existence of a 2-form 

To summarise we have shown that the embedding condition \eq{mc1} determines a
tensor multiplet on the $N=2,d=6$ worldsurface and that the components
of this multiplet satisfy their equations of motion. As we have seen,
these are somewhat complicated which is related to the fact that the
5-brane resembles a Dirichlet brane in some respects. As we pointed out
in \cite{hs}, the difference between type I and type II embeddings from
a geometrical point of view is that for the former there is an adapted
basis of the odd tangent bundle of the target space which splits into
components tangent and normal to the worldsurface whereas this is not so
for type II. In other words, $E_{\a}$ and $E_{\a'}$ are not related to
$E_{\ua}$ by a $Spin(1,10)$ matrix. The failure of adaptivity is due to
the presence of the tensor field $h$ and is a signal of
Dirac-Born-Infeld type behaviour; precisely this type of geometrical
structure also occurs for Dirichlet branes considered as
superembeddings. 

We note also that the geometry of the worldsurface is induced in the sense 
that the components of the worldsurface supervielbein can be expressed in 
terms of target space fields and the embedding matrix $\del_M Z^{\unM}$, 
up to gauge transformations.

The embedding condition \eq{mc1} leads directly to the equations of motion.
This is not surprising in view of the fact that it is not known how to
construct an off-shell version of the tensor multiplet. However, modulo
difficulties with self-duality, one might hope to find a `component'
action, in other words a Green-Schwarz type action. This is a very interesting 
question and is currently under study. 

In the case of Dirichlet branes, which form 
an interesting class of type II branes, Green-Schwarz type 
$\kappa$ invariant actions have been recently found in \cite{d1,d2}. It would be 
interesting to understand how the superembedding formalism is related to the 
Green-Schwarz formalism for Dirichlet and other type II branes. 
For type I branes, $\k$-symmetry is related to odd worldsurface diffeomorphisms 
as was first pointed out in \cite{stv}. 
In fact, for all branes, under an infinitesimal worldsurface diffeomorphism 
$\d Z^M=-V^M$ the variation of the embedding expressed in a preferred frame 
basis is
\beq
\d Z^{\unA}\equiv \d Z^{\unM} E_{\unM}{}^{\unA}= v^A E_A{}^{\unA}\ .
\eeq
For an odd transformation ($v^a=0$) one has
\beq
\d z^{\una}=0\qquad\d z^{\ua}= v^{\a} E_{\a}{}^{\ua}\ .
\eeq
The vanishing of the even variation $\d Z^{\una}$ is typical of $\k$-symmetry and
follows from the basic embedding condition \eq{mc1}. 

Finally, we mention the fact that we have assumed that the target space
geometry corresponds to on-shell $D=11$ supergravity. It may be possible
to derive this rather than take it as an input and there are indications
that this should be so. In particular we know that the requirement of
$\k$ symmetry for the 2-brane in the Green Schwarz formalism forces the
equations of motion \cite{bst}. However, the situation is more complicated here
in that one is trying to determine the form of the embedding matrix and
the dimension zero torsion on both the worldsurface and on the target
space from \eq{mc1}.

\pagebreak

\end{document}